\UseRawInputEncoding
\documentclass[prb,twocolumn,byrevtex,floatfix,superscriptaddress]{revtex4-1}

\usepackage{amssymb,amsmath,afterpage}
\usepackage{color,bm}
\usepackage[pdftex]{graphicx}
\pdfoutput=1
\newcommand*{\vect}[1]{\mathbf{#1}}

\begin{document}

\title{\boldmath Coexistence of antiferromagnetism and ferrimagnetism in adjacent honeycomb layers\unboldmath}

\author{D. Szaller$^\dagger$}
\altaffiliation{These authors contributed equally to this work}
\affiliation{Institute of Solid State Physics, TU Wien, 1040 Vienna, Austria}
\email[]{david.szaller@tuwien.ac.at}

\author{L. Prodan}
\altaffiliation{These authors contributed equally to this work}
\affiliation{Experimental Physics 5, Center for Electronic Correlations and Magnetism, Institute of Physics, University of Augsburg, D-86159, Augsburg, Germany}
\affiliation{Institute of Applied Physics, MD 2028, Chisinau, R. Moldova}

\author{K. Geirhos}
\affiliation{Experimental Physics 5, Center for Electronic Correlations and Magnetism, Institute of Physics, University of Augsburg, D-86159, Augsburg, Germany}

\author{V. Felea}
\affiliation{Experimental Physics 5, Center for Electronic Correlations and Magnetism, Institute of Physics, University of Augsburg, D-86159, Augsburg, Germany}
\affiliation{Institute of Applied Physics, MD 2028, Chisinau, R. Moldova}
\affiliation{Hochfeld-Magnetlabor Dresden (HLD-EMFL) and W\"{u}rzburg-Dresden Cluster of Excellence ct.qmat, Helmholtz-Zentrum Dresden-Rossendorf, 01328 Dresden, Germany}

\author{Y. Skourski}
\author{D. Gorbunov}
\author{T. F\"{o}rster}
\author{T. Helm}
\affiliation{Hochfeld-Magnetlabor Dresden (HLD-EMFL) and W\"{u}rzburg-Dresden Cluster of Excellence ct.qmat, Helmholtz-Zentrum Dresden-Rossendorf, 01328 Dresden, Germany}

\author{T. Nomura}
\affiliation{Hochfeld-Magnetlabor Dresden (HLD-EMFL) and W\"{u}rzburg-Dresden Cluster of Excellence ct.qmat, Helmholtz-Zentrum Dresden-Rossendorf, 01328 Dresden, Germany}
\affiliation{Institute for Solid State Physics, University of Tokyo, Kashiwa, Chiba 277-8581, Japan}

\author{A. Miyata}
\author{S. Zherlitsyn}
\affiliation{Hochfeld-Magnetlabor Dresden (HLD-EMFL) and W\"{u}rzburg-Dresden Cluster of Excellence ct.qmat, Helmholtz-Zentrum Dresden-Rossendorf, 01328 Dresden, Germany}

\author{J. Wosnitza}
\affiliation{Hochfeld-Magnetlabor Dresden (HLD-EMFL) and W\"{u}rzburg-Dresden Cluster of Excellence ct.qmat, Helmholtz-Zentrum Dresden-Rossendorf, 01328 Dresden, Germany}
\affiliation{Institut f\"{u}r Festk\"{o}rper- und Materialphysik, Technische Universit\"{a}t Dresden, 01062 Dresden, Germany}

\author{A. A. Tsirlin}
\affiliation{Experimental Physics 6, Center for Electronic Correlations and Magnetism, Institute of Physics, University of Augsburg, D-86159, Augsburg, Germany}

\author{V. Tsurkan}
\affiliation{Experimental Physics 5, Center for Electronic Correlations and Magnetism, Institute of Physics, University of Augsburg, D-86159, Augsburg, Germany}
\affiliation{Institute of Applied Physics, MD 2028, Chisinau, R. Moldova}

\author{I. K\'{e}zsm\'{a}rki}
\affiliation{Experimental Physics 5, Center for Electronic Correlations and Magnetism, Institute of Physics, University of Augsburg, D-86159, Augsburg, Germany}

\date{\today}

\begin{abstract}
Antiferromagnetic and ferro/ferrimagnetic orders are typically exclusive in nature, thus, their co-existence in atomic-scale proximity is expected only in heterostructures. Breaking this paradigm and broadening the range of unconventional magnetic states, we report here on an atomic-scale hybrid spin state, which is stabilized in three-dimensional crystals of the polar antiferromagnet Co$_2$Mo$_3$O$_8$ by magnetic fields applied perpendicular to the \emph{Co} honeycomb layers and possesses a spontaneous in-plane ferromagnetic moment. Our microscopic spin model, capturing the observed field dependence of the longitudinal and transverse magnetization as well as the magnetoelectric/elastic properties, reveals that this novel spin state is composed of an alternating stacking of antiferromagnetic and ferrimagnetic honeycomb layers. The strong intra-layer and the weak inter-layer exchange couplings together with competing anisotropies at octahedral and tetrahedral \emph{Co} sites are identified as the key ingredients to stabilize antiferromagnetic and ferrimagnetic layers in such a close proximity. We show that the proper balance of magnetic interactions can extend the stability range of this hybrid phase down to zero magnetic field. The possibility to realize a layer-by-layer stacking of such distinct spin orders via suitable combinations of microscopic interactions opens a new dimension towards the nanoscale engineering of magnetic states.
\end{abstract}

\maketitle

Exploring new magnetic phases with peculiar spin orders that can lead to unique material functionalities is a fundamental goal of spintronics~\cite{Jung16,Zhou18,Yan21}. While a great fraction of the spintronic devices is still based on ferromagnets and their intefaces/heterostructures with heavy metals~\cite{Zut04, Chap07, Manch19}, antiferromagnets progressively gain ground in such applications, as witnessed by the exponential growing field of antiferromagnetic (AFM) spintronics~\cite{Jung16, Balt18, Jung18}. Although ferromagnetism and antiferromagnetism are two mutually exclusive phases, the possibility to combine their best qualities triggered extensive research, leading to the realization of bilayers and heterostructures featuring a large exchange bias~\cite{Mei56,Radu08}. These artificial structures are typically composed of thin films of soft ferromagnets deposited on an AFM layer. Since the magnetization of the ferromagnetic layer is often unstable, being susceptible even to weak fluctuations of magnetic field and/or spin current, the exchange coupling to the anisotropic antiferromagnet is required to make the ferromagnetic (FM) state more robust for reliable device applications~\cite{Nog99,Chap07}. In fact, the exchange-bias effect is essential for current and future spin valve and MRAM technologies~\cite{Liu11,Park11,Bha17,Yug18,Lin19}.

In addition to metallic ferro- and antiferromagnets, recently magnetic insulators have also been considered as active elements in spintronic applications~\cite{Bea08,Bibes08,Chum14, Avc17,Jun18,Hor21}. Their main potential advantage, as compared to the metallic systems, is that their operation is based on electric-field control and not on current control, thus heat dissipation is considerably reduced. Furthermore, as opposed to metals, they are characterized by short-range magnetic interactions, i.e. lack the non-local Ruderman-Kittel-Kasuya-Yosida interactions~\cite{Kas56,Yos57}. Short-range interactions in insulating crystals could, in principle, lead to the realization of distinct magnetic orders even in atomic-scale proximity. This concept naturally requires materials that are composed of structural subunits, where the exchange interactions within these subunits are considerably stronger than between them.

As a proof of principle, we demonstrate the coexistence of antiferromagnetism and ferrimagnetism in adjacent honeycomb layers of Co$_2$Mo$_3$O$_8$. While this compound features a strict hierarchy of intra- and inter-layer exchange interactions, covalent and ionic chemical bondings between the neighbouring layers are still preserved, distinguishing this system from van der Waals magnets and heterostructures\cite{Sie21}. This unconventional hybrid phase emerges in external magnetic fields applied perpendicular to the layers as an intermediate state between the low-field easy-axis collinear AFM state and the high-field spin-flop phase. The latter two represent conventional magnetic phases with the same types of spin patterns in each honeycomb layer. As a further peculiar aspect of this intermediate phase, the ferromagnetic moment emerging at every second layer has a transverse component, that is even larger than the component parallel to the field. We show that competing magnetic anisotropies of chemically different \emph{Co} sites are indispensable for the layer-by-layer alternation of the magnetic state, in addition to the weak inter-layer coupling. In a simplistic spin model, we specify the general microscopic conditions, i.e. the ratio of exchange couplings and anisotropies, required to stabilize this hybrid phase without an external field.

\begin{figure}[t!]
\centering
\includegraphics[width=0.30\textwidth]{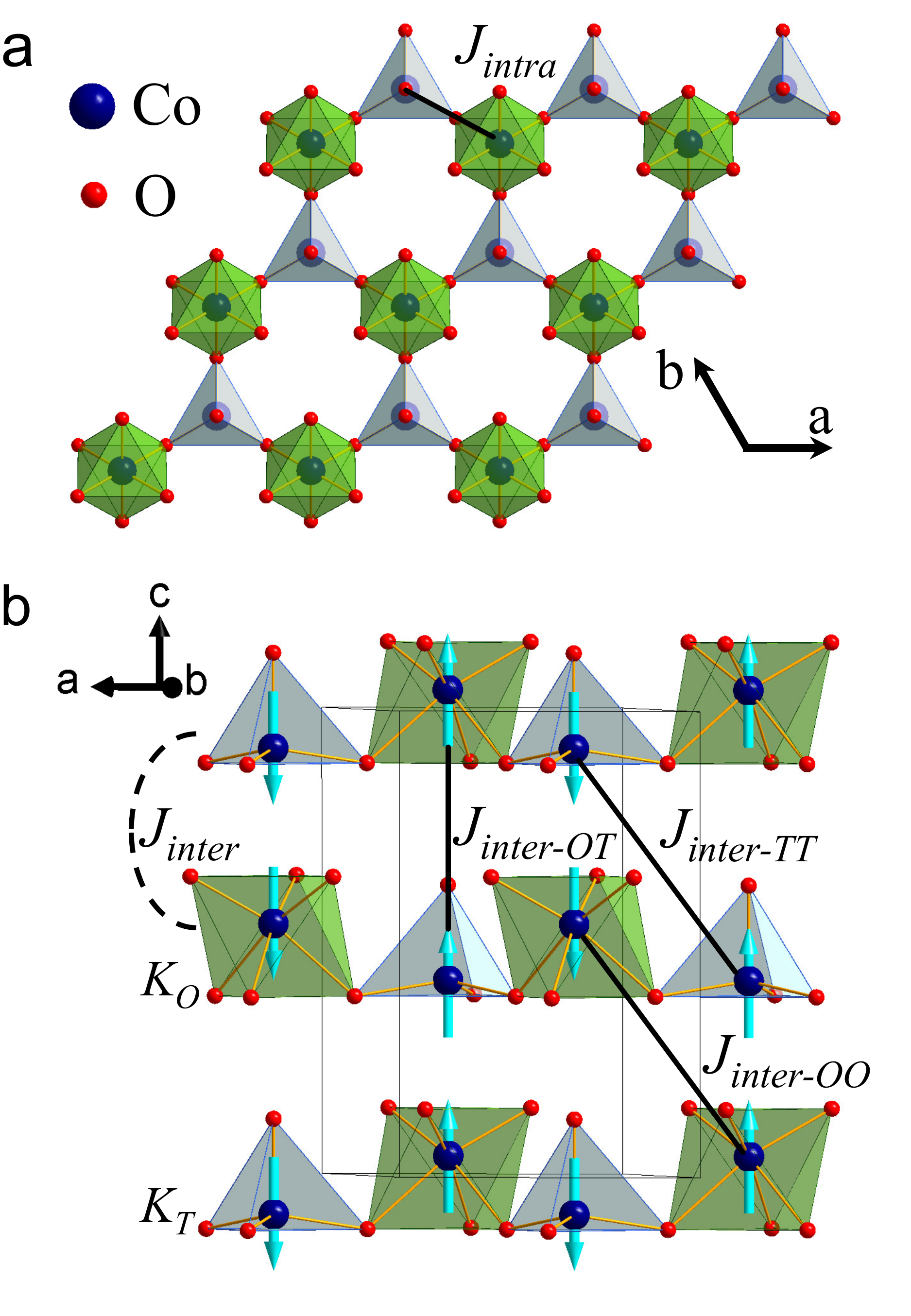}
\caption{\textbf{$\mid$ Crystal structure of Co$_2$Mo$_3$O$_8$.} (a) Co sites (blue spheres) and O positions (red spheres) in the $ab$ plane and (b) in the $ac$ plane including the spin pattern of the collinear AFM ground state \cite{Cze72}. Inter-spacer Mo ions are omitted. $J_{intra}$ stands for the intra-layer interaction between the octahedral and tetrahedral sites in the same honeycomb layer, while $J_{inter-OO}$, $J_{inter-TT}$, and $J_{inter-OT}$ are inter-plane interactions between octahedral sites, tetrahedral sites, and octahedral-to-tetrahedral sites, respectively. \emph{J}$_{inter}$ is an effective inter-layer coupling along the $c$ axis, capturing the effect of the three inter-layer exchange paths in a simplified scheme. Values of the magnetic interactions are listed up in Table I.}
 \label{fig_structure}
\end{figure}

Ternary transition-metal molybdenum oxides, \emph{M}$_2$Mo$_3$O$_8$ with \emph{M} = Mn, Fe, Co, and Ni, crystallize in a hexagonal polar structure (space group \emph{P6$_3$mc}). Their structure is a stack of honeycomb layers of magnetic \emph{M}$^{2+}$ ions along the $c$ axis, where adjacent \emph{M} sites have tetrahedral and octahedral coordinations in an alternating fashion both within the honeycomb layers and along the $c$ axis~\cite{McA83}, as shown in Fig.~\ref{fig_structure}. These materials exhibit fascinating phenomena, such as axion-type electrodynamics~\cite{Kur17}, high-temperature optical diode effect~\cite{Yu18}, vibronic excitations~\cite{Csi20}, doping induced ferrimagnetism~\cite{Kur15}, and precursor short-range magnetic order within the honeycomb layers~\cite{Res20}. Furthermore, these compounds were predicted to have topological spin-wave excitations \cite{Mor19}. According to their symmetry they are candidates to host N\'eel-type skyrmion lattices~\cite{Kez15}, in case ferro- or ferrimagnetic order can be stabilized, like in Mn$_2$Mo$_3$O$_8$ \cite{Sza20} and Zn-doped Fe$_2$Mo$_3$O$_8$~\cite{Kur15}.

These compounds undergo an AFM transition at temperatures between $T_N$=6-60~K~\cite{McA83,Kur15,Wan15,Kurum17,Mor19,Sza20}.
Not only $T_N$ but also the spin pattern in the ground state shows a strong variation with the transition metal ion. Without an external magnetic field, Co$_2$Mo$_3$O$_8$ and Fe$_2$Mo$_3$O$_8$ share a common easy-axis collinear AFM structure~\cite{Cze72,Ber75}, as depicted in Fig.~\ref{fig_structure}, where adjacent spins are arranged antiferromagnetically in the honeycomb layers and form ferromagnetic chains along the $c$ axis. This is a fully compensated AFM state, though in Fe$_2$Mo$_3$O$_8$ the individual honeycomb layers may possess a finite magnetization due to slightly different magnetic moments of the tetra- and octahedral \emph{Fe} sites~\cite{Cze72, Wan15}. In contrast, the net magnetization of all layers point along the same direction in Mn$_2$Mo$_3$O$_8$, forming an easy-axis ferrimagnet~\cite{Kurum17}. In Ni$_2$Mo$_3$O$_8$, a more complicated zig-zag AFM order has been reported~\cite{Mor19}. Such versatility of ground-state spin patterns, realized in the same crystal structure, implies a complex network of exchange paths and significant dependence of magnetic interactions on the orbital occupation, spin-orbit effects and spin-lattice coupling.

At first, we verify by susceptibility measurements that the layered crystal structure of Co$_2$Mo$_3$O$_8$ leads to strong magnetic anisotropies and confines the dominant exchange paths to individual honeycomb layers, as implied by the lack of dispersion of the magnon branches along the $c$ axis~\cite{CoMoO_magnon}. Indeed, we find a large difference between the susceptibility values obtained for fields parallel and perpendicular to the layers. Based on the temperature-dependent susceptibility curves, shown in Fig.~S1 of the supplement, we refined the basic set of magnetic interactions using quantum Monte Carlo simulations~\cite{Todo01}. The important aspects of the magnetic interactions revealed here are the dominance of intra-layer AFM exchange, $J_{intra}$= 14\,K, the weak inter-layer FM exchange, $J_{inter}$=-1.4\,K, the strong easy-axis anisotropy, $K_O$=14\,K, and the weak easy-plane anisotropy, $K_T$=-1.4K, of Co sites with octahedral and tetrahedral coordination, respectively. $J_{inter}$ refers to an overall effective inter-layer coupling, which has contributions from exchanges between octahedral sites ($J_{inter-OO}$) and tetrahedral sites ($J_{inter-TT}$) on adjacent layers according to $J_{inter}$=-($J_{inter-TT}$+$J_{inter-OO}$)/4. The different exchange paths and the single-ion anisotropies are depicted in Fig.~\ref{fig_structure}. This hierarchy of intra- and inter-layer exchanges as well as the presence of strong single-ion anisotropies, that are comparable to the dominant exchange coupling, are further supported by our DFT results. (The list of magnetic parameters is given in Table I; for details of the Monte Carlo and DFT calculations see the supplement).

\begin{table}[t!]
\caption{Comparison of exchange constants and single-ion anisotropies as determined by three different approaches: Quantum Monte Carlo simulation of temperature-dependent susceptibility, ab initio calculations (DFT) and mean-field spin model. All values are in kelvin units.}\label{tab1}
\begin{tabular}{|l|c|c|c|c|c|}
 \hline Monte Carlo & $J_{intra}$=14 & $J_{inter}$=-1.4  & $K_O$=14 & $K_T$=-1.4
\\ \hline DFT & $J_{intra}$=18 & $J_{inter}$=-2.2 & $K_O$=7.7 & $K_T$=2.3
\\ \hline Mean field & $J_{intra}$=12 & $J_{inter}$=-1.5 & $K_O$=26.8 & $K_T$=-1.73
\\ \hline
\end{tabular}
\end{table}

\begin{figure}[t!]
\centering
\hspace{-0.0\textwidth}\includegraphics[width=0.46\textwidth]{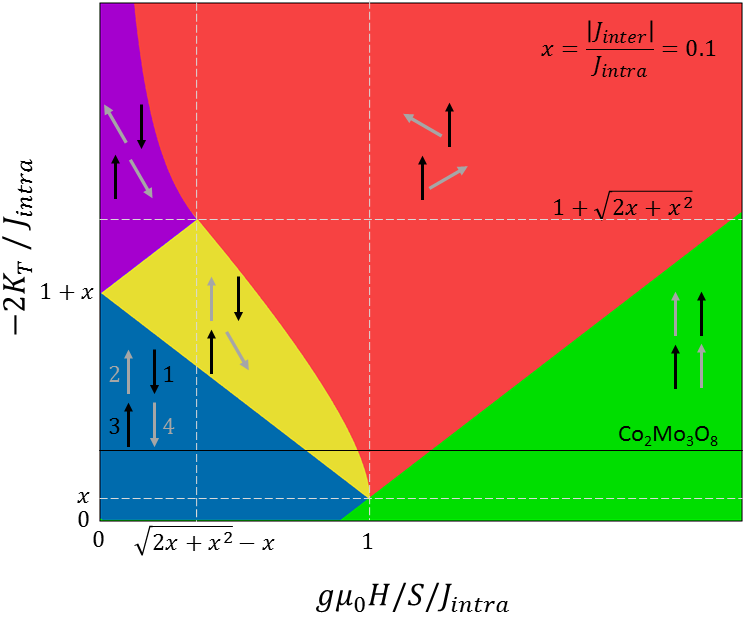}
\caption{\textbf{$\mid$ Magnetic phase diagram of the simplistic four-spin model.} Ground-state spin orders on the magnetic anisotropy ($K_T$) versus Zeeman energy plane for magnetic fields applied parallel to the vertical Ising spins (black arrows).
Spins indicated by gray arrows experience $K_T<0$ easy-plane anisotropy, favouring horizontal spin alignment. $J_{intra}>0$ is the AFM exchange between spins in the same horizontal "layer", while $J_{inter}=-xJ_{intra}$ is the ferromagnetic coupling between spins on adjacent "layers", as depicted in Fig. 1. For these calculations the value of $x=0.1$ was taken. Dashed horizontal and vertical lines indicate special values of  $(K_T)$ and the magnetic field ($H$), respectivley. The anisotropy value corresponding to the case of Co$_2$Mo$_3$O$_8$ is also indicated by horizontal black line. Purple, red, yellow, blue and green regions correspond to the tilted antiferromagnetic (TAF), symmetric spin-flop (SF), asymmetrically canted (ASC), collinear antiferromagnetic (AF), and field-polarized ferromagnetic (FM) phases, respectively.}
 \label{fig_K2H}
\end{figure}

To explore possible magnetic orders emerging from the magnetic interactions identified above, we constructed a simplified spin model~\cite{Pee19,Sza20}. We describe the four-spin unit cell of Co$_2$Mo$_3$O$_8$, by introducing Ising spins on the two octahedral \emph{Co} sites ($\vect{S}_1$ and $\vect{S}_3$), while spins at the tetrahedral sites ($\vect{S}_2$ and $\vect{S}_4$) are characterized by easy-plane anisotropy, $K_T<0$. The pairs in the same layers, $\vect{S}_1-\vect{S}_2$ and $\vect{S}_3-\vect{S}_4$, are coupled by a strong antiferromagnetic exchange $J_{intra}>0$, while the $\vect{S}_1-\vect{S}_4$ and $\vect{S}_2-\vect{S}_3$ pairs are connected by the weaker $J_{inter}<0$ ferromagnetic coupling. We studied the ground state of this model by varying $K_T$ and the strength of the magnetic field, applied along the Ising spins. The phase diagram in Fig.~\ref{fig_K2H}, obtained for $J_{inter}=-J_{intra}/10$, reveals five distinct phases. At most four of them can be realized for a given set of parameters by varying the magnetic field. While this choice of the inter- and intra-layer exchanges is realistic for Co$_2$Mo$_3$O$_8$ according to our Monte Carlo and DFT calculations (see Table I and Fig.~\ref{fig_K2H}), we approximate the magnetic state of octahedral sites by Ising spins, by formally setting $K_O$ to infinity. In each magnetic phase, depicted in Fig.~\ref{fig_K2H} and discussed below, we illustrate the magnetic order schematically. Black and gray arrows represent the vertical Ising spins and the spins with horizontal easy-plane anisotropy, respectively.

If $K_T$ is sufficiently weak, $2\lvert K_T\rvert<\lvert J_{inter}\rvert$, the low-field collinear antiferromagnetic phase (blue AFM region) directly transforms into the field-polarized state (green FM region) at the saturation field $g\mu_0 H_S/S = J_{intra}+J_{inter}-2K_T$. Thus, for $H<H_S$ the magnetization is zero and it jumps to the saturation value at $H_S$.

With increasing $K_T$, two intermediate states appear between the AFM and FM regions, where either one or both of the tetrahedral $\vect{S}_2$ and $\vect{S}_4$ spins are canted towards their easy plane. In the asymmetric canted state (yellow, ASC) the equivalency of these two spins is broken, since the exchange field and the external magnetic field are parallel for one spin, while they are antiparallel for the other. Correspondingly, only the spin pointing opposite to the field in the collinear AFM state can gain Zeeman energy by canting towards the easy plane, which leads further gain in anisotropy energy. The other tetrahedral spin stays co-aligned with the Ising spins. Consequently, the ASC phase exhibits a spontaneous transverse magnetization. Interestingly, for the canting-angle range of $0<\vartheta<\pi/2$ the magnetization component transverse to the field exceeds the component along the field, $M_{trans}/M_{long}=\sin\vartheta/(1-\cos\vartheta)$, thus, a significant magnetic torque is expected. On the contrary, in the spin-flop phase (red SF region) the transverse magnetization vanishes due to the opposite (symmetric) canting of the two tetrahedral spins. Please note that in our simplistic model the ASC phase is the realization of the desired hybrid state with alternating AFM and ferrimagnetic layers.

For even stronger easy-plane anisotropy, $\lvert K_T\rvert>(J_{intra}+\lvert J_{inter}\rvert)/2$, a tilted antiferromagnetic phase (purple TAF region) becomes the zero-field ground state, where the tetrahedral spins stay antiparallel with each other yet they are not co-aligned with the Ising spins anymore, but tilted away to their easy plane instead. In finite fields, the magnetization of this phase has finite components both along and perpendicular to the field.

\begin{figure*}[t!]
\includegraphics[width=0.8\textwidth]{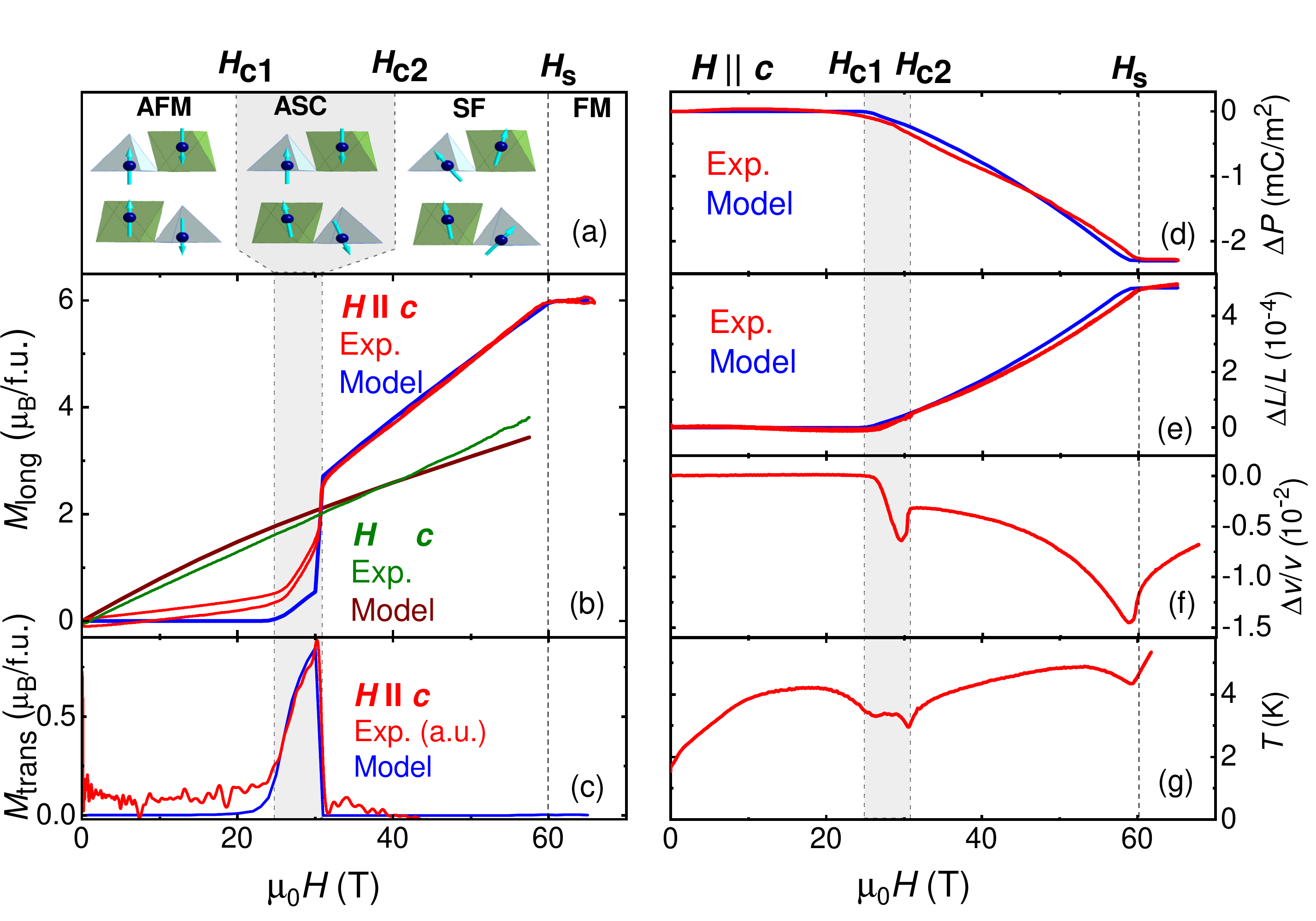}
 \caption{\textbf{$\mid$ Sequence of magnetic phases in Co$_2$Mo$_3$O$_8$.}(a) Spin configurations in the collinear antiferromagnetic (AFM), asymmetric canted (ASC), symmetric spin flop (SF) and spin-polarized ferromagnetic (FM) phases. Color coding for the octahedral and tetrahedral sites is the same as in Fig.~\ref{fig_structure}. (b) Magnetization measured in magnetic fields parallel (red line) and perpendicular (green line) to the $c$ axis at 1.5\,K, together with the corresponding curves (blue and brown) obtained from the mean-field model. (c) Transverse magnetization as measured (red line) and as predicted by the mean-field model (blue line). The measured data are plotted in arbitrary units, since the calibration of the magnetic torque was not possible in pulsed fields. Field dependence of (d) the magnetically induced polarization $\Delta P$, (e) the relative change of the sample length $\Delta L/L$, (f) the relative change of the sound velocity $\Delta v/v$, and (g) the change of the sample temperature (magnetocaloric effect) at 1.5\,K with fields parallel to the $c$ axis. Blue lines in panels (d) and (e) show the results of the mean-field model. In each panel, vertical dashed lines mark the critical fields $H_{c1}$ and $H_{c2}$ and the saturation field $H_{s}$, while the region of the intermediate ASC phase is indicated by a grey area.}
\label{fig_fielddep}
\end{figure*}

In the following, we test the predictions of our spin model on Co$_2$Mo$_3$O$_8$, which is expected to offer an ideal laboratory for exploring a layer-by-layer alternation of AFM and ferrimagnetic states. First, we study the evolution of its ground state by high-field magnetometry. Fig.~\ref{fig_fielddep}(b) shows the magnetization measured in pulsed fields applied parallel and perpendicular to the $c$ axis at 1.5\,K. For $H\parallel c$, the magnetization shows only a weak increase up to the first critical field, $H_{c1}$=25\,T, above which it grows rapidly, leading to a kink-like feature at $H_{c1}$. With further increasing the field, a sudden jump of the magnetization to nearly half of the saturation value is observed at the second critical field, $H_{c2}$=30\,T. Then, via a linear increase, the saturation to the spin-only moment of 6\,$\mu_B$/f.u. is reached at $H_s$ =60\,T. For $H\perp c$, the magnetization shows a nearly linear increase over the whole field range with a slope approximately half as large as for $H_{c2}<H<H_s$ in the perpendicular geometry and without any sign of metamagnetic transitions. This sequence of phase transitions, with two intermediate phases between the zero-field AFM state and the field-polarized FM state, is in line with the prediction of our spin model, as clear from Fig.~\ref{fig_K2H}.

In order to show that the phase between $H_{c1}$ and $H_{c2}$ has the ASC spin texture with finite transverse moment, we also performed high-field torque measurements. Fig.~\ref{fig_fielddep}(c) displays the transverse magnetization data in arbitrary units, since we could not properly calibrate our cantilever torque magnetometry setup due to technical limitations in pulsed fields. Nevertheless, these experimental data confirm that only this phase exhibits finite transverse magnetization, in agreement with our simplistic model. In fact, as demonstrated in Figs.~\ref{fig_fielddep}(a)-(c), a fully quantitative description of the magnetization data can be achieved by refining the parameters of this spin model. (The only feature not captured by our mean-field model is the finite longitudinal susceptibility of the collinear AFM state below $H_{c1}$, which likely has an inherently quantum origin and arises from spin-stretching modes of the S=3/2 spins~\cite{Pen12}.) As the main step of this refinement, the Ising spins at the octahedral sites have been replaced by spins with strong but finite easy-axis anisotropy, $K_O=26.8$\,K. The full parameter set best reproducing the magnetization data is given in Table I.

As described before, the four sublattices in our model correspond to two neighboring Co sites within a single honeycomb layer (one with tetrahedral and one with octahedral coordination) and the two adjacent Co sites in the next layer, as sketched in Figs.~\ref{fig_structure} and \ref{fig_fielddep}(a). Therefore, the material-specific calculations on Co$_2$Mo$_3$O$_8$ reveal that the phase between $H_{c1}$ and $H_{c2}$ is composed of alternating layers of distinct spin patterns. Namely, every second layer hosts collinear AFM order, similar to the zero-field ground state, while spins in the intermediate layers are canted away from the $c$ axis. Moreover, the spin canting at the tetrahedral sites is larger than the canting of spins to the opposite direction at the octahedral sites, due to the weak easy-plane and strong easy-axis anisotropy of the respective sites. The alternating stacking of collinear AFM and canted ferrimagnetic layers is reproduced by the simulation also for larger system sizes along the $c$ axis, excluding finite-size artifacts. Above $H_{c2}$, spins in all layers become canted, thereby completing the transition to the conventional spin-flop (SF) phase. Due to their opposing anisotropies, the octahedral and tetrahedral spins exhibit different canting, but the spontaneous magnetic moment perpendicular to the $c$ axis is cancelled. The spin arrangement in the different phases is visualized in Fig.~\ref{fig_fielddep}(a).

\begin{figure}[t!]
\hspace{-0.008\textwidth}\includegraphics[width=0.49\textwidth]{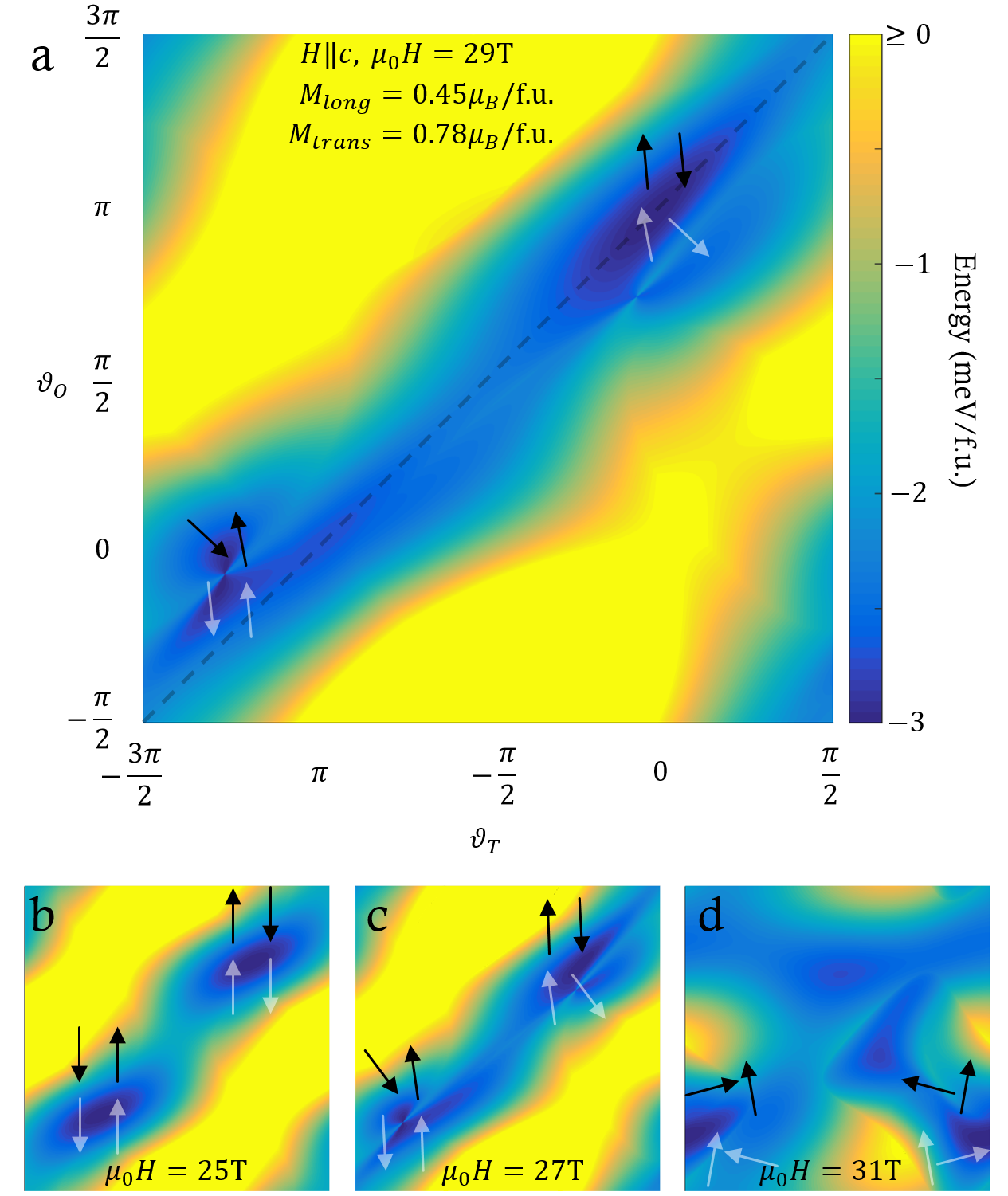}
 \caption{\textbf{$\mid$ Stability of magnetic phases in Co$_2$Mo$_3$O$_8$.} Heatmap plots of the energy, obtained for spin configurations with magnetization constrained to the experimental value, on the $\vartheta_O$--$\vartheta_T$ plane. $\vartheta_O$ and $\vartheta_T$ correspond to the azimuthal angles of the octahedrally ($S_1$) and tetrahedrally coordinated ($S_2$) spins in the same layer, respetively. (For more details, see the main text.) The angular ranges are common for each map. Energies are obtained from the mean-field spin model for magnetic field applied along the $c$ axis. The energy minima correspond to the collinear antiferromagnetic phase at 25\,T (b), the asymmetric canted phase at 27\,T (c) and 29\,T (a), and  symmetric spin-flop phase at 31\,T (d). (a) The diagonal dashed line in panel (a) corresponds to the collinear antiferromagnetic allignement of $S_1$ and $S_2$. At the two global minimuma, the orientations of $S_1$ and $S_2$ are illustrated by black arrows, while spins on the next layer are depicted by gray arrows. In these cartoons the vertical axis corresponds to the $c$ direction.
 \label{fig_heatmap}}
\end{figure}

Next, we investigate the nature of the metamagnetic transitions induced by fields along the $c$ axis by calculating the energy of the four-spin magnetic unit cell for a large set of spin configurations. To reduce the number of independent (spin) variables, we constrained this analysis to spin configurations having the same total magnetization as observed in the experiments. Specifically, the spins in one layer ($\vect{S}_1$ and $\vect{S}_2$) were fixed, while the orientation of the spins in the other layer ($\vect{S}_3$ and $\vect{S}_4$) was varied to minimize the energy. The heatmaps in Fig.~\ref{fig_heatmap} show the energy landscape obtained at selected magnetic fields on the plane of the azimuthal angles, $\vartheta_O$ and $\vartheta_T$, spanned respectively by $\vect{S}_1$ and $\vect{S}_2$ and the $c$ axis. Here the diagonal, connecting the bottom left and top right corners, corresponds to the collinear AFM order of $\vect{S}_1$ and $\vect{S}_2$. In $H<H_{c1}$ magnetic fields, as presented in Fig.~\ref{fig_heatmap}(b), there are two equivalent energy minima, corresponding to anti-phase domains of the collinear AFM state. In the ASC phase [Figs.~\ref{fig_heatmap}(a) and (c)], the two minimuma again represent anti-phase domains, where adjacent layers show asymmetric canting and host collinear AFM spin texture in reversed order. The continuous evolution of the energy landscape accross $H_{c1}$ [compare Figs.~\ref{fig_heatmap}(b) and (c)] indicates a second-order transition, while the abrupt change at $H_{c2}$ [see Figs.~\ref{fig_heatmap}(c) and (d)] implies a first-order transition.

Besides the sequence of spin patterns, we also studied structural changes induced via the spin-lattice coupling throughout the metamagnetic transitions. Figs.~\ref{fig_fielddep}(e) and (f) show that the magnetostriction, $\Delta$L/L, and the magnetically-induced polarization, $\Delta P$, both measured along the $c$ axis, follow the same trend. In the collinear AFM both are constant. Above $H_{c1}$ the unit cell expands along the $c$ axis and $\Delta$L/L reaches a value of 5$\times$10$^{-4}$ at saturation, while $\Delta P$ levels off at approximately -2350\,$\mu$C/m$^2$, close to the magnitude reported for the giant-magnetoelectric compound Fe$_2$Mo$_3$O$_8$ \cite{Kur15,Wan15}. These changes indicate a considerable magnetoelastic coupling. Based on the field-dependent spin configurations obtained from our model, the magnetostriction as well as the polarization change were reproduced by utilizing their proportionality to the spin-spin correlations~\cite{Joh17,Kim15}. We found intra-layer $\langle S_OS_T\rangle$-like correlations to be dominant, implying a strong distance dependence of the intra-layer exchange, $J_{intra}$. Indeed, DFT structure relaxations for different magnetic states suggest the shortening of the lattice constant within the layers and the simultaneous increase in the inter-layer spacing upon going from AFM to FM order. This leads to a 1.8\,\% increase in $J_{intra}$, whereas $J_{inter}$ remains unchanged within the accuracy of our calculations. (For more details, see the supplement.)

To gain further insight into the magnetoelastic properties, we measured the field dependence of the sound velocity, $\Delta v/v$, with propagation vector and polarization along and perpendicular to the $c$ axis, respectively. The data are shown in Fig.~\ref{fig_fielddep}(f). In analogy with the magnetostriction, the sound velocity stays constant up to $H_{c1}$. However, as the most prominent feature, it shows a large dip in the ASC state. Magnon-phonon coupling, e.g. via the exchange-striction mechanism, often results in a renormalization of the acoustic properties at magnetic transitions~\cite{Lut05}. Since acoustic magnon modes play a dominant role in such processes, the strong reduction of the sound velocity in the ASC state likely indicates the emergence of soft spin-wave excitations. Indeed, due to its transverse ferromagnetic moment, the ASC phase breaks the rotational symmetry of the spin-spin interactions about the $c$ axis and consequently turns one of the optical magnon branches to an acoustic one. The emergence of such a zero-energy Goldstone mode in the center of the Brillouin zone, also confirmed by our spin model, is likely responsible for the dip of the field-dependent sound velocity in the ASC state. In the ordinary spin-flop phase above $H_{c2}$, the rotational symmetry is still broken by the in-plane moments, hence the sound velocity further varies with an overall change of $\Delta v/v$=-1.5\,\% up to the saturation field. In the field-polarized FM state the sound velocity shows a sudden upturn, as expected due to the lack of a Goldstone mode. The above scenario is also supported by the adiabatic temperature change in pulsed magnetic fields, the so-called magnetocaloric effect (MCE), shown in Fig.~\ref{fig_fielddep}(g). Besides sharp peaks at the critical fields, the most striking feature in the MCE curve is the drop in the ASC state. Since MCE originates from the field-derivative of the specific heat and/or the magnetization, this dip again implies the emergence of soft magnons.

\begin{figure}[t!]
\centering
\includegraphics[width=0.44\textwidth]{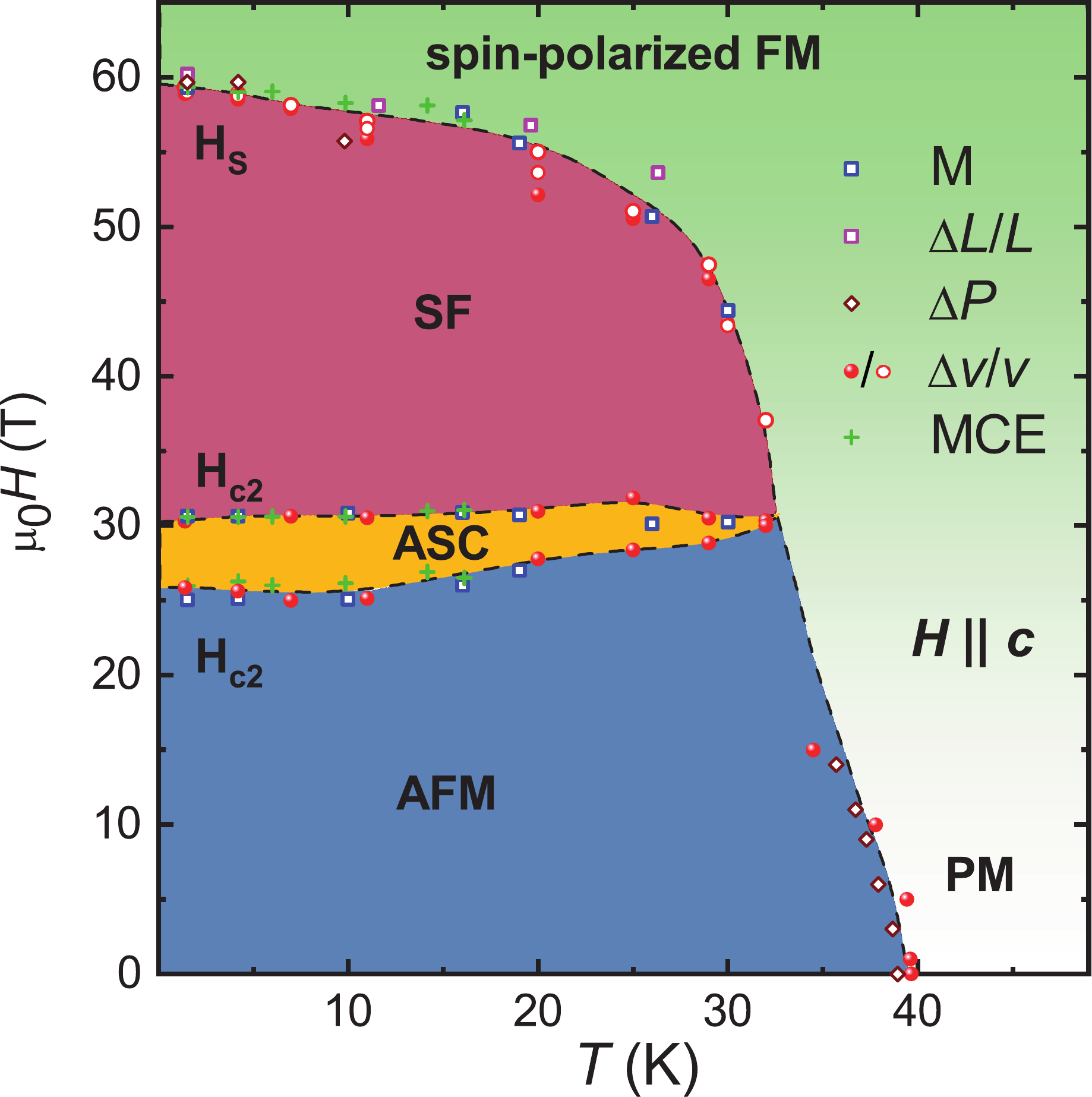}
\caption{\textbf{$\mid$ Magnetic phase diagram of Co$_2$Mo$_3$O$_8$ for fields along the \emph{c}-axis,} comprising antiferromagnetic (AFM, blue), asymmetric canted (ASC, mustard), spin flop (SF, wine), spin-polarized ferromagnetic (FM, green), and paramagnetic (PM, white) phases. The critical fields $H_{c1}$ and $H_{c2}$ as well as the saturation field $H_{s}$ are determined from magnetization (M), magnetostriction ($\Delta L/L$) and magnetocaloric effect (MCE) measurements in pulsed fields, from polarization ($\Delta P$) and ultrasound velocity ($\Delta v/v$) measurements in static and pulsed fields. For ultrasound experiments both the velocity and the attenuation are plotted with full and open red symbols, respectively, for longitudinal and transverse acoustic waves. }
\label{fig_phases}
\end{figure}

Finally, we investigated the thermal stability range of the ASC phase by repeating the field-dependent measurements at various temperatures between 1.5-50\,K. As clear from Fig.~\ref{fig_phases}, the ASC phase persist up to approximately 33\,K, i.e. it keeps separating the collinear AFM phase and the SF phase up to the paramagnetic state.

In conclusion, we observed a sequence of metamagnetic transitions in Co$_2$Mo$_3$O$_8$ with magnetically weakly coupled honeycomb layers. We found that, due to a delicate interplay of strong intra-layer and weak inter-layer couplings with competing easy-axis and easy-plane anisotropies, the collinear AFM order is only partially destabilized at the lowest critical field: the collinear AFM state is preserved in every second honeycomb layer, while a ferrimagnetic spin configuration emerges in the intermediate layers. Based on a mean-field spin model, supplemented by ab initio calculations and quantum Monte Carlo simulations, we achieve a consistent refinement of the full set of magnetic interactions relevant in this prototype honeycomb antiferromagnet, and quantitatively describe the magnetic and magnetoelectric/elastic properties throughout the four magnetic phases. We also show that the hybrid phase with AFM and ferrimagnetic layers can extend to zero field if the easy-plane anisotropy is sufficiently strong. The concept to realize distinct spin states in atomic-scale proximity, achieved in this honeycomb system with quasi-2 dimensional magnetic interactions, can be naturally extended to van der Waals magnets, to quasi-1 dimensional systems and crystals with molecular-magnet-like building blocks.

\section{Methods}

Polycrystalline Co$_2$Mo$_3$O$_8$ was prepared by solid-state reactions of binary oxides CoO (99.999\,\%) and MoO$_2$ (99\,\%) in evacuated quartz ampoules by repeated sintering at 1000\,$^\circ$C. Single crystals were grown by a chemical transport reaction method between 950 and 900\,$^\circ$C using anhydrous TeCl$_4$ as source of the transport agent. Crystals up to a size of 7\,mm were obtained after four weeks of transport. Phase purity of the samples was confirmed by x-ray powder diffraction on crushed single crystals. Magnetization, magnetoelectric, magnetostriction, magnetocaloric and sound velocity measurements were performed in static and pulsed fields up to 65\,T at low temperatures on single crystals of Co$_2$Mo$_3$O$_8$.

Magnetic interactions and magnetic anisotropy were obtained by relativistic DFT band-structure calculations performed using the FPLO~\cite{Koe99} and VASP~\cite{Kre96} codes utilizing the Perde-Burke-Ernzerhof flavor of the exchange-correlation potential~\cite{Per96}. Correlation effects in the Co $3d$ shell were taken into account on the mean-field level via DFT+$U$ with the on-site Coulomb repulsion parameter $U_d$=5\,eV, Hund's exchange $J_d$=1\,eV, and double-counting correction in the atomic limit~\cite{Wu05}. Magnetic parameters of the spin Hamiltonian
\begin{equation}
\mathcal{H} = \sum_{<i,j>}{J}_{ij}\vect{S}_{i}\vect{S}_{j} - \sum_{i}{K}_{i}({S}_{i}^z)^2
\end{equation}
were obtained using a mapping procedure~\cite{Xia11}. Here, $J_{ij}$ is the exchange interaction between lattice sites $i$ and $j$, $K_i$ is the single-ion anisotropy at site $i$ and S=3/2. In addition to DFT, classical mean-field and quantum Monte Carlo simulations were used to refine the exchange and anisotropy parameters of Eq.~(1). In the mean-field model, the spin configurations minimizing the energy were numerically determined for the whole experimentally studied field range. While a larger set of magnetic interactions were identified in DFT ($J_{intra}$, $J_{inter-OO}$, $J_{inter-TT}$, $J_{inter-OT}$, $K_O$, $K_T$), only the minimal set ($J_{intra}$, $J_{inter}$, $K_O$, $K_T$), necessary to reproduce the observed temperature- and field-dependent magnetic properties, were kept in the Monte-Carlo and mean-field calculations.

The mean-field results were obtained by finding the minimal energy of the four-spin cluster,
\begin{widetext}
\begin{multline}
\mathcal{H}_{MF} = 3{J}_{intra}\left(\vect{S}_{O1}\vect{S}_{T1}+\vect{S}_{O2}\vect{S}_{T2}\right) + 2{J}_{inter}\left(\vect{S}_{O1}\vect{S}_{T2}+\vect{S}_{T1}\vect{S}_{O2}\right)\\ - {K}_{O}\left(({S}_{O1}^z)^2 + ({S}_{O2}^z)^2\right) - {K}_{T}\left(({S}_{T1}^z)^2 + ({S}_{T2}^z)^2\right)
-g\mu_0\mu_\textrm{B} \vect{H}\left(\vect{S}_{O1}+\vect{S}_{T1}+\vect{S}_{O2}+\vect{S}_{T2}\right),
\end{multline}
\end{widetext}
where $\vect{S}_{O1}$ and $\vect{S}_{O2}$ are the classical spin vectors corresponding to the Co ions in octahedral environment in the odd and even $ab$-plane layers of the crystal, respectively. Similarly, $\vect{S}_{T1}$ and $\vect{S}_{T2}$ represent the spins of the tetrahedrally coordinated Co ions. Integer multiplicators of the exchange terms correspond to the coordination numbers. The g-factor, vacuum permeability, Bohr magneton and magnetic field are denoted by $g$, $\mu_\textrm{B}$, $\mu_\textrm{0}$ and $\vect{H}$, respectively.

The Hamiltonian of the simplistic four-spin model (Fig.~\ref{fig_K2H})  was
\begin{widetext}
\begin{equation}
\mathcal{H}_{simpl} = {J}_{1}\left({S}_{1}{S}_{2}^z+{S}_{3}{S}_{4}^z\right) + {J}_{2}\left({S}_{1}{S}_{4}^z+{S}_{2}^z{S}_{3}\right) - {K}_{2}\left(({S}_{2}^z)^2 + ({S}_{4}^z)^2\right)
-g\mu_0\mu_\textrm{B} {H}^z\left({S}_{1}+{S}_{2}^z+{S}_{3}+{S}_{4}^z\right),
\end{equation}
\end{widetext}
where $S_1=\pm S$ and $S_3=\pm S$ are Ising spins.

\textbf{Acknowledgements} We acknowledge S.-W. Cheong, S. Bordacs, J. Deisenhofer, O. Zaharko and A. Pimenov for fruitful discussions. This work was supported by the Deutsche Forschungsgemeinschaft (DFG) through Transregional Research Collaboration TRR 80 (Augsburg, Munich, and Stuttgart), SFB 1143, and the excellence cluster ct.qmat (EXC 2147, Project ID 39085490), and by the BMBF via DAAD (Project No. 57457940), as well as by the project ANCD 20.80009.5007.19 (Moldova). D.S. acknowledges the support of the FWF Austrian Science Fund grants I 2816-N27 and TAI 334-N. We acknowledge support by HLD at HZDR, member of the European Magnetic Field Laboratory (EMFL).

\textbf{Author Contributions} D.S. and L.P. contributed equally to this work. L.P., K.G., V.T. performed the measurements in static magnetic fields. V.F., Y.S., D.G., T.F., T.N., A.M., S.Z., J.W. performed the experiments in pulsed magnetic fields. V.T. analysed the data. L.P., V.T. synthesized and characterized the single crystals. T.H. fabricated the lamella for the cantilever torque magnetometry measurements. A.T. performed the Monte Carlo and the ab initio calculations. D.S. performed the mean-field calculations. I.K. wrote the manuscript with contributions from D.S., V.T. and A.T. I.K. planned the project.

\textbf{Additional information} The authors declare no competing financial interests. Supplementary information accompanies this paper. Correspondence and requests for materials should be addressed to D.S. and I.K.

\end{document}